%Paper: hep-th/9211094
%From: feher@avzw02.physik.uni-bonn.de (Laszlo Feher)
%Date: Sun, 22 Nov 92 11:36:48 GMT

%**** This is PlainTEX file, no macro needed ****

\magnification=\magstep1
\hsize=16.0 true cm
\vsize=22.0 true cm
\baselineskip=14 true pt
\parindent=3 true pc
\parskip=3 true pt
\tolerance=5000

\def\A{{\cal A}}
\def\g{{\cal G}}

\def\tr{{\rm tr\,}}
\def\La{\Lambda}
\def\ellg{\ell(gl_n)}

\def\pa{\partial}

\def\res{{\rm res\,}}

\font \ninerm                 = cmr9
\font \ninesl                =cmsl9

\centerline{
{\bf GENERALIZED DRINFELD-SOKOLOV HIERARCHIES AND
${\cal W}$-ALGEBRAS}${}$\footnote*{{\ninerm
Talk given at the NSERC-CAP Workshop on
{\ninesl Quantum Groups, Integrable Models and Statistical Systems},
Kingston, Canada, July 13-17 1992.}}
}
\bigskip

\centerline{L. Feh\'er${}$\footnote{**}{{\ninerm On leave
 from Bolyai Institute of Szeged University, H-6720 Szeged, Hungary.}}
}
\bigskip
\centerline{\it Laboratoire de physique nucl\'eaire}
\centerline{\it
Universit\'e  de Montr\'eal, C.~P. 6128, Montr\'eal, Canada H3C 3J7}
\bigskip

\centerline{\bf ABSTRACT}
\bigskip
{\baselineskip=12 true pt
\ninerm
\narrower\smallskip\noindent
We review the construction of Drinfeld-Sokolov type hierarchies and
classical ${\cal W}$-algebras in a Hamiltonian symmetry reduction framework.
We describe the list of graded regular elements
in the Heisenberg subalgebras of the
nontwisted loop algebra $\ellg$
and deal with the associated hierarchies.
We exhibit an $sl_2$ embedding
for each reduction of a Kac-Moody Poisson bracket
algebra to a ${\cal W}$-algebra
of gauge invariant differential polynomials.\smallskip}
\bigskip
\baselineskip=14 true pt
\noindent
{\bf 1. Review of the Drinfeld-Sokolov Construction}
\medskip
In this talk I wish to describe some recent results
on the construction of KdV type hierarchies and classical
${\cal W}$-algebras.
(Proofs and further details can be found in [1], [2].)
First I review the relevant aspects of the
Drinfeld-Sokolov (DS) construction of KdV type hierarchies [3]
and the corresponding ${\cal W}$-algebras
concentrating on the simplest case.
I shall raise some questions concerning the possible
generalizations,
which will be (partially) answered later in the talk.

As explained in detail in [1], the  DS
construction can be naturally understood in the framework
of the Hamiltonian
Adler-Kostant-Symes approach to integrable systems (e.g.~[4]).
The hierarchy results
  from a local symmetry
reduction of the commuting family of Hamiltonian systems generated
by the ${\rm ad}^*$-invariant Hamiltonians on the dual
$\A^*$ of a Lie algebra $\A$ of the form
$$
\A = \ell({\cal G}):= {\cal G}\otimes {\bf C}[\lambda,\lambda^{-1}],
\eqno(1.1)$$
where ${\cal G}$ itself is a centrally extended loop algebra.
The space ${\cal A}^*$ carries the family of compatible
R Lie-Poisson brackets induced by the classical r-matrices
$R_k\in {\rm End} (\A)$ given by $R_k:= (P_+-P_-)\circ \lambda^k$,
where $P_\pm\in {\rm End}(\cal A )$ project onto
the subalgebras $\A_\pm$ containing positive and negative powers
of the spectral parameter $\lambda$, respectively, (see [5]).

For simplicity, let us concentrate on the case when
$\g={\widetilde {gl_n}}^\wedge$, the central extension of
the algebra of smooth loops in $gl_n$, i.e.,
$\g=\{\,(X,a)\,\vert\,X: S^1\to gl_n\,,\, a\in {\bf C}\,\}$
with the Lie bracket
$$
[(X,a),(Y,b)]=\left( XY-YX\,,\,\int_0^{2\pi} dx\,{\rm tr}X'(x)Y(x) \right)\ .
\eqno(1.2)
$$
(Convention: The periodic space variable parametrizing $S^1$
is denoted by $x\in [0,2\pi]$ and tilde
signifies ``loops in $x$''. For any space $V$, we set
$\ell(V):=V\otimes {\bf C}[\lambda,\lambda^{-1}]$.)
In the usual way, the dual space $\A^*$ (or a dense subspace thereof) is
represented as the space of first order matrix differential
operators $\cal L$ of the form
$$
{\cal L}=(e\partial_x + \mu(x)),
\eqno(1.3)
$$
where
$\mu(x)=\sum \mu_i(x) \lambda^i$
is a mapping from $S^1$ into
$\ell(gl_n):= gl_n\otimes {\bf C}[\lambda,\lambda^{-1}]$,
and $e=\sum e_i\lambda^i$
is an element of ${\bf C}[\lambda, \lambda^{-1}]$.
The  ${\rm ad}^*$-invariant functions are generated by the
invariants (eigenvalues) of the monodromy matrix $T(\lambda)$ of
${\cal L}$.

A crucial r\^ole in the consruction is played by the
``DS matrix'' $\La_n$ given by
$$
\Lambda_n
=\left[\matrix{
0&1&0&\cdots&0\cr
\vdots&0&1&\ddots&\vdots\cr
\vdots&{}&\ddots&\ddots&0\cr
0&{}&{}&\ddots&1\cr
\lambda &0&\cdots&\cdots&0\cr}\right].
\eqno(1.4)$$
This is a {\it regular element} of $\ellg$,
that is, we have
$$
\ellg = {\rm Ker\,}({\rm ad\,}\Lambda)
+{\rm Im\,}({\rm ad\,}\Lambda)\,,
\qquad
{\rm Ker\,}({\rm ad\,}\Lambda)
:\ \hbox{abelian subalgebra}\,,
\eqno(1.5)$$
for $\La =\La_n$.
In fact, ${\rm Ker\,}({\rm ad\,}\Lambda_n)$
is the principal Heisenberg subalgebra of $\ellg$
(it acquires the central extension in $\ellg^\wedge$).
Further,  $\La_n$ has grade $1$ in the principal grading
of $\ellg$.
A grading of $\ellg$
can be defined by the eigenspaces of a derivation
$d_{N,H} :\ellg\rightarrow \ellg$ of the form
$$
d_{N,H}:=N\lambda {d\over d\lambda} + {\rm ad\,} H\,,
\eqno(1.6)$$
where $N$ is an integer and $H\in gl_n$  is
diagonalizable.
The principal grading is obtained by taking
$N:=n$ and $H:=H_n$, where
$$
H_n:={1\over 2}{\rm diag\,}[(n-1),(n-3),\ldots,-(n-3),-(n-1)].
\eqno(1.7)$$
We shall need the decomposition
$$
gl_n=gl_n^- + gl_n^0 + gl_n^+
\eqno(1.8)$$
induced by the eigenvalues of ${\rm ad\,}H_n$,
where the summands are the subalgebras of (strictly) lower triangular,
diagonal
and upper triangular matrices,
and also
the constant matrices $C_0$ and $C_1$ defined by writing
$\La_n$ in the form
$$
\Lambda_n := C_0 + \lambda C_1\,.
\eqno(1.9)$$

The DS construction starts by restricting to the subspace
${\cal M}\subset \ell({\widetilde{gl_n}}^\wedge)^*$ consisting of operators of
the form
$$
{\cal L}=\partial +J+\lambda C_1\,,
\qquad
J : S^1\to gl_n\ .
\eqno(1.10)$$
This is a Poisson subspace with respect to
two out of the  infinitely many $R$ Lie-Poisson brackets on
$\ell({\widetilde{gl_n}}^\wedge)^*$.
The corresponding
compatible Poisson brackets on ${\cal M}$  are given by
$$
\{\varphi, \psi\}_1 (J) =
-\int_{S^1}\tr C_1[{{\delta \varphi}\over {\delta J}} ,
{{\delta \psi}\over {\delta J}} ]\,,
\eqno (1.11)
$$
$$
\{\varphi, \psi\}_2 (J) =
\int_{S^1}\tr \left(J [{{\delta \varphi}\over {\delta J}} ,
{{\delta \psi}\over {\delta J}}  ] +
\left({{\delta \varphi}\over {\delta J}}\right)^\prime
{{\delta \psi}\over {\delta J}}\right)\,,
\eqno (1.12)
$$
where ${{\delta \varphi}\over {\delta J}}$
(resp.~${{\delta \psi}\over {\delta J}}$)
is the functional derivative of
the function $\varphi$  (resp.~$\psi$)  on ${\cal M}$.
The monodromy invariants of ${\cal L}$ provide Hamiltonians
on $\cal M$ that form a commutative family
with respect to both Poisson brackets.

It follows from
$$
[gl_n^-\,,\,C_1]=0
\eqno(1.13)$$
that the group ${\cal N}$ of transformations
$$
e^f : {\cal L}\mapsto e^{f}{\cal L}e^{-f}\,,
\qquad\hbox{with}\qquad  f : S^1\to  gl_n^-\ ,
\eqno(1.14)$$
is a  symmetry group of the commuting
family of bihamiltonian systems carried by ${\cal M}$.
Indeed, these transformations preserve the two Poisson structures
and the monodromy invariants.
The KdV type hierarchy results from a
{\it symmetry reduction} defined by using
${\cal N}$ in such a way as  to ensure the {\it locality}
of the reduced system.
Concretely, one considers
the following two step reduction process.
First, one restricts the system to the ``constrained manifold''
${\cal M}_c\subset {\cal M}$, defined as the set of
${\cal L}$'s of the following special form:
$$
{\cal L}=\partial + (j+C_0)+\lambda C_1 =\partial +j+\Lambda_n\,,
\qquad
j: S^1\to (gl_n^-+gl_n^0)\ .
\eqno(1.15)$$
Second, one factorizes this constrained manifold by the
symmetry group ${\cal N}$, defining the reduced phase space
$$
{\cal M}_{\rm red}={\cal M}_c/{\cal N}\ .
\eqno(1.16)$$
In other words,  one factorizes out the
``gauge transformations'' generated by ${\cal N}$ by declaring
that only the ${\cal N}$-invariant functions of ${\cal L}$
are physical.
This reduction has the following nice features:

\noindent
\item{i)}
  The  eigenvalues of the monodromy matrix of ${\cal L}\in {\cal M}_c$
   can be computed by a recursive, algebraic procedure and
   thus they give commuting {\it local} Hamiltonians.

\noindent
\item{ii)}
    The compatible Poisson brackets on ${\cal M}$ induce compatible
    Poisson brackets on ${\cal M}_{\rm red}$.

\noindent
\item{iii)}
     The  gauge orbits in ${\cal M}_c$  allow for global, differential
     polynomial gauge sections, which
     give rise to
     free generating sets for the
     gauge invariant differential polynomials in $j$.

\noindent
\item{iv)}
The gauge invariant
differential polynomials form a classical
${\cal W}$-algebra under the second Poisson bracket.
\smallskip

The monodromy invariants of a first order matrix differential
operator are in general {\it nonlocal} objects.
Statement i) is a consequence of the facts
that $\La_n\in \ellg$ is a {\it graded regular element of
nonzero grade}, and the grades in $j$ (1.15)
are smaller than the grade of $\La_n$.
The point is that by substituting the ansatz
$$
\Psi(x) = (I+Z(x))e^{F(x)}(I+Z(0))^{-1}\Psi(0)\,,
\eqno(1.17)$$
with $F(x)\in {\rm Ker\,}({\rm ad\,}\Lambda_n)$,
$Z(x)\in {\rm Im\,}({\rm ad\,}\Lambda_n)$,
into the linear problem ${\cal L}\Psi =0$,
one can determine both $Z$ and $F$ by quadrature
by using the grading together with the
decomposition (1.5). One can also easily
diagonalize the monodromy matrix
$T=\Psi(2\pi)\Psi(0)^{-1}$.
The resulting ${\rm ad}^*$-invariant Hamiltonians are {\it local};
i.e., are given by
integrals of local densities formed
  from the components of $j$ and their derivatives.

To construct new local hierarchies,
it would be important
to explore the possible constraints
on the form of a first order matrix differential
operator under which the monodromy invariants are local.
It is known that one can
associate such constraints to any
positive graded regular element of any affine Lie algebra
(see [6]).
However, the list of the inequivalent graded regular elements of
the affine Lie algebras seems to be unknown.

Statement ii) means that the compatible Poisson brackets carried
by ${\cal M}$ can be consistently restricted to the {\it gauge invariant}
functions on ${\cal M}_c$.
This follows from the Dirac theory of reduction by constraints.
By choosing some basis $\{\gamma_i\}$ of $gl_n^-$, the
constraints defining ${\cal M}_c\subset {\cal M}$ read
$$
\phi_i(x)=0\,
\qquad\hbox{where}\qquad
\phi_i(x):= {\rm tr\,}\gamma_i (J(x)-C_0)\,.
\eqno(1.18)$$
It is easy to verify  that they are {\it first class}
with respect to  any of the compatible Poisson brackets on ${\cal M}$.
The $\phi_i(x)$ are in fact
the generating densities
of the ${\cal N}$ symmetry transformations
with respect to the second Poisson bracket (1.12).
Therefore the Dirac theory tells us that we should
factorize the constrained manifold by these transformations.
The second Poisson bracket closes on the gauge
invariant functions, which can be identified with the functions
on ${\cal M}_{\rm red}$.
Thus we obtain an induced Poisson bracket on the factor space
(the Dirac bracket).
On the other hand, the $\phi_i$ do not generate
any transformations on ${\cal M}$  under the first Poisson bracket (1.11);
i.e.,~they are ``Casimir functions''.
Therefore the first Poisson bracket can in principle already be restricted to
${\cal M}_c$ without any factorization by ${\cal N}$.
Then ${\cal N}$ becomes a group of Poisson (canonical)
transformations with respect to the
restricted bracket, which can further be reduced to a Poisson bracket
on the invariant functions.
In this way, we naturally obtain two induced Poisson brackets
on ${\cal M}_{\rm red}$ from those on ${\cal M}$,
and the induced Poisson brackets are compatible because the
original brackets (1.11,12) were compatible.

The gauges appearing in statement iii) are defined as follows [3].
Consider a direct sum decomposition
$$
(gl_n^-+gl_n^0)=[C_0,gl_n^-]+V,
\eqno(1.19{\rm a})$$
where the linear space $V$ is graded by
eigenvalues of ${\rm ad\,}H_n$ (i.e., $[H_n,V]\subset V$).
Then the subspace of ${\cal M}_c$ consisting of operators of the form
$$
{\cal L}_V=\pa + j_V +\La_n,
\qquad
j_V: S^1\to V,
\eqno(1.19{\rm b})$$
defines a global gauge section.
As proven in [3], a general element
${\cal L}\in {\cal M}_c$, given by (1.15),
can be brough to  this gauge  by a unique gauge
transformation $e^f\in {\cal N}$
and $f$ is a {\it differential polynomial} in $j$.
It follows that the components of $j_V$, when considered as functions
on ${\cal M}_c$, give a basis for the gauge invariant differential
polynomials  in $j$, which thus form a
{\it freely generated} differential ring.
In [2] we gave a fairly general sufficient condition
for the existence of this type of gauges (which we call
``DS gauges'')
in reductions by first class constraints.
It should perhaps be noted that DS gauges are {\it not} available
for the vast majority of reductions.

Let us now deal with statement iv).
Note first that the differential polynomial
$$
L_{H_n} := {1\over 2} {\rm tr} (J^2) + {\rm tr\,}(H_n J^\prime )
\eqno(1.20)$$
satisfies the Virasoro algebra under
the second Poisson bracket, and its restriction
to ${\cal M}_c$ is gauge invariant.
Since it contains this Virasoro density,
the second Poisson bracket algebra of the gauge invariant
differential polynomials is an extended conformal
algebra.
Set $M_0:=H_n$, $M_+:=C_0$ and choose $M_-\in gl_n$ so that
the $sl_2$ relations
$$
[M_0,M_\pm]=\pm M_\pm\,,\qquad[M_+,M_-]=2M_0
\eqno(1.21)$$
hold.
Consider the particular DS gauge belonging to
$$
V:= {\rm Ker\,}({\rm ad}_{M_-})\,.
\eqno(1.22)$$
A graded basis of this $V$ is given by the matrices
$(M_-)^k$ with $k=0,\ldots, (n-1)$.
It can be shown [7] that, with the exception of the $M_-$ component,
the gauge invariant differential polynomials corresponding to
the components  of the gauge fixed current $j_V$
are in this case all primary fields (conformal tensors) with
respect to the conformal action generated by $L_{H_n}$.
These primary fields and $L_{H_n}$ together
form a basis (free generating set) for the
gauge invariant differential polynomials.
This means that the extended conformal algebra of the
gauge invariant differential polynomials is indeed a classical
$\cal W$-algebra.

The gauge transformations are generated by the constraints through
the second Poisson bracket (1.12), which
can be recognized as a ``Kac-Moody Poisson bracket'' (namely,
the Lie-Poisson bracket corresponding to the affine Lie algebra
$\widetilde{gl_n}^\wedge$).
The property that
the gauge invariant differential polynomials
form a ${\cal W}$-algebra concerns only the
reduction of the Kac-Moody (KM) Poisson bracket algebra,
and is largely independent from other features of the hierarchy.
We shall return to the generalizations of this
${\rm KM} \longrightarrow {\cal W}$ reduction  at
the end of the talk.
\bigskip
\noindent
{\bf 2. Graded Regular Elements in Heisenberg
Subalgebras of $\ellg$}
\medskip
Drinfeld and Sokolov [3] associated integrable hierarchies
to the grade $1$ generators of the principal Heisenberg
subalgebras of the loop algebras,
given by $\La_n$ (1.4) in the case of $\ellg$.
The fact that these are {\it graded regular} elements
is crucial for obtaining the {\it local} monodromy invariants
giving the Hamiltonians of the hierarchies.
Recently, it has been proposed by
De Groot {\it et al} [6] to construct new integrable
hierarchies (and new ${\cal W}$-algebras)
by using {\it any} positive, graded regular
element of {\it any } Heisenberg subalgebra
of a loop algebra in a ``generalized DS construction''.
Roughly speaking,  a set of constraints and a gauge group
was associated to each graded regular element.
Clearly, the actual content of this proposal depends
on the supply of graded regular elements,
which has not been investigated in [6].
The inequivalent graded Heisenberg subalgebras of the
affine Lie algebras were
classified by Kac and Peterson [8]
and an explicit description of them was
worked out by ten Kroode and van de Leur in Refs.~[9], [10].
By using this explicit description, it is not hard to
obtain the list of the graded regular elements by inspection.

The graded Heisenberg subalgebras of $\ellg$
are classified by the {\it partitions} of $n$ in the
following way [9].
Let a partition of $n$ be given by
$$
n=n_1+n_2+\cdots +n_k\ ,
\qquad \hbox{where}\qquad
n_1\geq n_2\geq \cdots \geq n_k\geq 1\ .
\eqno(2.1)$$
The corresponding Heisenberg subalgebra
consists of the $n\times n$ ``block-diagonal''
matrices $\La$ of the form
$$
\La=\left[\matrix{ y_1 \La_{n_1}^{l_1}&{}&{}&{}\cr
              {}&y_2\La_{n_2}^{l_2}&{}&{}\cr
              {}&{}&\ddots&{}\cr
             {}&{}&{}&y_k\Lambda^{l_k}_{n_k}\cr}\right]\ ,
\eqno(2.2)$$
where the $l_i$ ($i=1,2,\dots,k$) are arbitrary integers, the $y_i$
are arbitrary  numbers, and $\La_{n_i}$ is
the $n_i\times n_i$ DS matrix, cf.~(1.4).
This maximal abelian subalgebra of $\ellg$ is invariant
under a grading operator $d_{N,H}$ of the form (1.6)
with $N$ and $H$ determined by the partition [9].
An  element $\Lambda$ in (2.2) is {\it regular}
-- gives rise to a decomposition of type (1.5) --
if ${\rm Ker}({\rm ad\,} \Lambda)\subset \ellg$
is the Heisenberg subalgebra (and not a larger space).

The simplest case is that of the {\it homogeneous} Heisenberg
subalgebra,
 belonging to the partition $n=1+1+\cdots + 1$.
In this case the  grading operator is
$\lambda{d\over d\lambda}$ and the graded regular elements are
of the form
$\Lambda = \lambda^k {\rm diag}[y_1,y_2,\ldots,y_n]$,
where $y_i\neq y_j$ for $i\neq j$ and $k$
is arbitrary integer.
The other extreme case  is that of
the {\it principal} Heisenberg subalgebra,
when $n$ is ``not partitioned at all''.
On account of $\Lambda_n^{l+mn}=\lambda^m \Lambda_n^l$,
the generator
$\Lambda_n^{l+mn}$ (of grade $(l+mn)$) is regular
if and only if $\Lambda_n^l$ is regular.
The DS matrix $\La_n$  itself is  regular
since its eigenvalues are the $n$ {\it distinct} $n$th-roots of
$\lambda$.
  From this one verifies,
by inspecting the eigenvalues of $\Lambda_n^l$,
that  for $1\leq l \leq (n-1)$ $\Lambda_n^l$
is regular if and only if $n$ and $l$ are relatively prime.
For the general case, we have the following result [1].
\medskip
\noindent
{\bf Theorem 1.} {\it Graded regular elements exist only in those
Heisenberg subalgebras of $\ellg$ which belong to the
partitions of type
$$
n=p r = \overbrace{r+\cdots +r}^{p\;\rm times}\ ,
\qquad\hbox{or}\qquad
n=pr+1=\overbrace{r+\cdots +r}^{p\;\rm times}+1\,.
\eqno(2.3)$$
In the equal block case $n=pr$ with $r>1$,
the graded regular elements are of the form
$$
\La=\lambda^m
\left[\matrix{ y_1 \La_{r}^{l}&{}&{}&{}\cr
              {}&y_2\La_{r}^{l}&{}&{}\cr
              {}&{}&\ddots&{}\cr
             {}&{}&{}&y_p\La^{l}_{r}\cr}\right] \, ,
\eqno(2.4)$$
where
$$
1\leq l\leq (r-1),\qquad
y_i\neq 0,\qquad y_i^r\neq y_j^r \qquad  i,j=1,\ldots,p\,,
\quad i\neq j\,,
$$
with $l$ relatively prime to $r$ and $m$ any integer.
The element $\Lambda$ is of grade $(l+mr)$, where
the grading operator $d_{N,H}$ is given by
(1.6) with $N=r$ and}
$$
H={\rm diag}[\overbrace{H_r,H_r,\ldots,H_r}^{p\;\rm times}]\,.
\eqno(2.5)$$
{\it In the equal-blocks-plus-singlet case $n=pr+1$,
the graded regular elements are those
$n\times n$ matrices which
contain an $(n-1)\times (n-1)$ block of the form
given by (2.4) in the ``top-left corner''
and an arbitrary entry in the
``lower-right corner''.
The relevant grading operator is
given by (1.6) with $N=r$,
$$
H={\rm diag}[\overbrace{H_r,H_r,\ldots,H_r}^{p\;\rm times},0]
\qquad\qquad \hbox{\it if $r$ is odd}\,;
\eqno(2.6{\rm a})$$
and with $N=2r$,}
$$
H={\rm diag}[\overbrace{2H_r,2H_r,\ldots,2H_r}^{p\;\rm times},0]
\qquad\qquad\hbox{\it if $r$ is even}\,.
\eqno(2.6{\rm b})$$
\smallskip

It would be interesting to know the list of
graded regular elements and associated
integrable systems for all loop algebras based
on the simple Lie algebras.
The above result, which is  of course also valid
in the case of $\ell(sl_n)$,
makes it clear that graded regular elements
exist only in some ``exceptional'' Heisenberg
subalgebras in general.
Some knew integrable systems can
presumably be obtained by applying the DS construction
to each graded regular element,
or in some cases one will recover known systems
and gain a better understanding of them in this way.
\bigskip
\noindent
{\bf 3.  Matrix Gelfand-Dickey Hierarchy from DS Reduction}
\medskip
In [1] we gave a detailed analysis of
the DS reduction based on a grade $1$ regular element
of the Heisenberg subalgebra of $\ellg$ defined by
a partition of type $n=pr$ with $r>1$,
generalizing the $r=n$ case described in [3].
After a reordering of the basis,
our grade $1$ regular element, $\La_{r,p}$, can be written as
$$
\Lambda_{r,p}=\Lambda_r \otimes D =
\left[\matrix{
0&D&0&\cdots&0\cr
\vdots&0&D&\ddots&\vdots\cr
\vdots&{}&\ddots&\ddots&0\cr
0&{}&{}&\ddots&D\cr
\lambda D&0&\cdots&\cdots&0\cr}\right]\,,
\eqno(3.1)$$
where the $p\times p$ matrix
$D:={\rm diag\,}(y_1,y_2,\ldots,y_p)$ is
such that $D^r$ has distinct, non-zero eigenvalues (cf.~(2.4)).
In this basis, the grading operator is given by
$d_{N,H}= r \lambda{d\over d\lambda} + {\rm ad\,} H$
with
$$
H=H_r\otimes 1_p=
{\rm diag\ }[\,j 1_p\,,\, (j-1)1_p\,,\,\ldots \,,\,-(j-1)1_p\,,\, -j 1_p\,]\,,
\quad j={(r-1)\over 2}.
\eqno(3.2)$$
In particular, this $H$ naturally gives every $n\times n$
matrix a block structure, with $p\times p$ blocks.
The DS reduction is set up quite similarly as in the $p=1$ case.
We introduce the matrices $C_0$ and $C_1$
through the equality $\La_{r,p}:=C_0+\lambda C_1$,
and define the spaces ${\cal M}$, ${\cal M}_c$,
and the gauge group ${\cal N}$ simply by
substititing ``block-triangular'' for ``triangular''
everywhere in the original definitions.
The following statements
identify the reduced system as
the $p\times p$ matrix
version of the well-known (e.g.~[11]) Gelfand-Dickey $r$-KdV hierarchy.

First, the reduced space,
${\cal M}_{\rm red}={\cal M}_c/{\cal N}$, is the space
of ``matrix Lax operators'' of the form
$$
L=(-D)^{-r} \partial^r + u_1 \partial^{r-1}+\ldots
+u_{r-1}\pa +u_r\ ,
\eqno(3.3)$$
where the $u_i$ are smooth, $p\times p$ matrix valued
functions on $S^1$.
Second, the Poisson brackets on ${\cal M}_{\rm red}$
induced by the reduction
are the two compatible matrix Gelfand-Dickey Poisson brackets,
given by the well-known formulae
$$
\eqalignno{
\{\varphi,\psi\}^{(1)}(L)&=
\int_{S^1}\tr{\rm res\,}\left(L [Y_-,X_-]\right)\,,
&(3.4)\cr
\{\varphi,\psi\}^{(2)}(L)&=
\int_{S^1}\tr{\rm res\,}\left(YL(XL)_+ -LY(LX)_+\right)\,,
&(3.5)\cr}
$$
where $X:=\nabla_L\varphi$, $Y:=\nabla_L\psi$
are the gradients of the functions $\varphi$, $\psi$ on ${\cal M}_{\rm red}$.
(Similarly as in the scalar case,
these gradients are pseudo-differential
operators and the subindex $\pm$ refers to the splitting
of the space of -- now $p\times p$ matrix -- pseudo-differential
operators into the sum of the subspaces of pure differential
 and integral operators, contaning positive and
negative powers of $\pa$.)
The second Poisson bracket algebra
qualifies as a classical ${\cal W}$-algebra.
Third, the Hamiltonians of the hierarchy,
resulting from the monodromy invariants of ${\cal L}\in {\cal M}_c$,
allow for the following description in terms of $L$:
Diagonalize $L$ by a recursive procedure; i.e.,
determine a $p\times p$ {\it diagonal} pseudo-differential operator
$\hat L=(-D)^{-r}\pa^r+\sum_{i=1}^\infty a_i \pa^{r-i}$ such that
$L=g\hat L g^{-1}$, where $g$ is of the form
$g=I+\sum_{i=1}^\infty g_i \pa^{-i}$
with $g_i(x+2\pi)=g_i(x)$.
A natural generating set for the Hamiltonians
of the hierarchy is
obtained by integrating the componentwise
residues of the fractional (including integral) powers $\hat L$.
More precisely, the list of Hamiltonians reads
$$
{\cal H}_{0,i}=(-1)^r\int_{S^1} \left(D^r u_1\right)_{ii}\,,
\,\qquad
{\cal H}_{k,i}={r\over k}\int_{S^1} \res \left({\hat
L}^{k/r}\right)_{ii}\,,
\eqno(3.6)$$
where $i=1,\ldots,p$ and $k=1,2,\ldots$ is arbitrary.
These Hamiltonians satisfy ``bihamiltonian ladder''
relations,
$\{L \,,\,{\cal H}_{k,i}\}^{(2)}=\{L \,,\,{\cal H}_{k+r,i}\}^{(1)}$.
The number of independent bihamiltonian ladders is
$n-1$ since one has  $\sum_{i=1}^p {\cal H}_{mr,i}=0$
for any $m=1,2,\ldots$, which is simply a consequence of
the fact that $L$ is a purely differential operator.

The above description of the reduced system
generalizes the result proven
by Drinfeld and Sokolov in [3] for the scalar case $p=1$.
The proofs given in [1]  use their methods, but
at the same time introduce some conceptual simplifications
(at least to our taste).
The simplifications arise from the fact  that we work entirely
within the Hamiltonian Adler-Kostant-Symes approach.
In this framework the existence of the compatible Poisson
structures and commuting Hamiltonians is clear from
the very beginning of the construction
and the only problem is to describe them in terms of reduced
variables as explicitly and nicely as possible.

The main difference between the
$p\times p$ matrix and the $p=1$ scalar case
is that computing the Hamiltonians
in the former case requires the diagonalization
of $L$.
The analogues of those Hamiltonians which are obtained from
the {\it integral} powers of the diagonalized Lax operator
$\hat L$ do not exist in the scalar case.
The other Hamiltonians can also be
expressed as integrals of trace-residues of independent
fractional powers of $L$, without diagonalization.

The KdV type hierarchies based on matrix Lax operators
of the type (3.3) have been investigated before in
refs.~[12-14],
where the additional assumption
was made that the diagonal part of $u_1$ vanishes.
We verified that setting $[u_1]_{\rm diag}=0$
is consistent with the equations of the
hierarchy resulting from the DS reduction
and in fact corresponds to an additional Hamiltonian
symmetry reduction.

Let us further comment on the relationship
between the hierarchies and ${\cal W}$-algebras.
It is known ([15], [2]) that one can  associate
a classical ${\cal W}$-algebra to
every $sl_2$ subalgebra of $gl_n$.
The  ${\cal W}$-algebra arising in the above
corresponds to the
$sl_2$ subalgebra under which the defining
representation of $gl_n$ decomposes into $p$ copies of
the $r$-dimensional $sl_2$ irreducible representation.
The other case
in which a graded regular element exists in the
Heisenberg subalgebra is the case of the partition
$n=pr+1$.
By taking an arbitrary regular element of minimal positive
grade it may be verified that the generalized DS reduction
proposed in [6] leads to a ${\cal W}$-algebra
which is again equal to one of those studied in [15], [2].
(We note in passing that it is not clear to us whether
the reductions belonging
to regular elements of higher grade are related
to ${\cal W}$-algebras or not.)
Both the $sl_2$ subalgebras of $gl_n$
and the Heisenberg subalgebras of $\ell(gl_n)$
are classified by the partitions of $n$.
It is unclear whether there is a general relationship
between all ${\cal W}$-algebras associated to
$sl_2$ embeddings and KdV type hierarchies
or not, since there is a ${\cal W}$-algebra
for any partition, but graded regular elements
exist only in exceptional cases.
It is also worth noting that,
in all cases, it is easy to construct families
of ``first Poisson structures'' compatible
with the ``second one'' giving the ${\cal W}$-algebra.
This fact however does not automatically imply
the existence of a corresponding local hierarchy.
\bigskip
\noindent
{\bf 4.  Are $sl_2$ Embeddings Necessary for ${\cal W}$-Algebras?}
\medskip
Consider a finite dimensional Lie algebra $G$
with an $\rm ad$-invariant,
nondegenerate scalar product
$\langle\ ,\ \rangle$.
Denote by ${\cal K}$  the space of $G$-valued
smooth, periodic functions, ${\cal K}:=\{\,J\,\vert\, J: S^1\to G\,\}$,
and let ${\cal K}$ carry the ``KM Poisson bracket algebra'':
$$
\{
\langle u , J(x)\rangle \,,\,\langle v , J(y)\rangle \}
=\langle [u,v],J(x)\rangle\delta (x-y)
-\langle u,v\rangle \delta^{\prime}(x-y)\,,
\eqno(4.1)$$
where $u$ and $v$ are arbitrary generators of $G$.
Choose a subalgebra $\Gamma\subset G$
(with basis $\{\gamma_i\}$)
and an element
$C_0\in G$ in such a way that the following constraints
$$
\phi_i(x)=0\,
\qquad\hbox{where}\qquad
\phi_i(x):= \langle\gamma_i\,,\, J(x)-C_0\rangle\,,
\eqno(4.2)$$
are {\it first class}.
The corresponding constrained manifold ${\cal K}_c\subset {\cal K}$
consists of ``currents'' of the form
$$
J=C_0 +j
\qquad\hbox{with}\qquad
j: S^1\to \Gamma^\perp\,.
\eqno(4.3)$$
The first class constraints $\phi_i$ generate
gauge transformations on ${\cal K}_c$ and we are interested
in the {\it gauge invariant differential polynomials}
in $j$.
We would like to describe and classify
the constraints for which the gauge invariant differential
polynomials form a classical ${\cal W}$-algebra
``similarly as in the standard DS case''.
It has been recently recognized [15, 2] that one can find
at least one such reduction for every $sl_2$ subalgebra of $G$,
by generalizing the standard case in a
rather straightforward way.
Thus it is natural to ask whether the presence
of an $sl_2$ embedding is necessary in all ``nice'' cases.
We do not have a complete classification of the nice cases yet,
but, under the assumptions given below, we can answer this
latter question in the positive.

Let ${\cal R}$ be the set of gauge invariant differential
polynomials in $j$.
This set is obviously closed with respect to
linear combination,
ordinary multplication and application of $\partial$.
We express this by saying that ${\cal R}$ is a
{\it differential ring}.
First of all, we assume that ${\cal R}$ is
{\it freely generated} on
$m:=({\rm dim\,} G -2{\rm dim\,}\Gamma )$
gauge invariant differential polynomials.
In other words, there exist generators $W^a\in {\cal R}$
($a=1, \ldots, m$) such that any element $W\in {\cal R}$
can be expressed in a unique way as a differential polynomial
in the $W^a$'s.
It follows that, upon imposing the first class
constraints, the KM Poisson bracket algebra
induces a Poisson bracket algebra of the form
$$
\{ W^b(x)\,,\, W^c(y)\}^* =
\sum_k  P^{b,c}_k(x) \delta^{(k)}(x-y)\,,
\eqno(4.4)$$
where the $P_k^{b,c}$ entering the finite sum on the right hand
side are uniquely determined
differential polynomials in the basis $W^a$ ($a=1,\ldots,m$).
We assume that this induced Poisson bracket gives
${\cal R}$ the structure of a {\it classical ${\cal W}$-algebra}.
By definition, this means that
it is possible to choose a {\it primary field
basis} in ${\cal R}$, that is, a basis  such that
$W^1$ is a Virasoro density and
$W^a$ is a conformal primary field for $a=2,\ldots, m$.

We now make a further assumption, which is more technical
 than the above, though we consider it still rather natural.
Namely, we assume that  the  primary field basis is such that
$W^1$ is equal to
$$
L_H={1\over 2}\langle J ,J \rangle +\langle H,J^{\prime}\rangle\,,
\eqno(4.5)$$
for {\it some} digonalizable element $H\in G$.
We can translate the fact that $L_H$
is gauge invariant into the relations
$$
[H,\Gamma]\subset \Gamma\,,
\qquad
H\in \Gamma^\perp\,,
\qquad
[H,C_0]=C_0\,.
\eqno(4.6)$$
Furthermore, we assume that the DS type gauges
are available by using this $H$ as grading operator.
The meaning of the latter assumption is the following
(cf.~Eq.~(1.19)).
Take any graded linear space $V$ defining a direct sum
decomposition
$$
\Gamma^\perp = [C_0,\Gamma] +V\,,
\eqno(4.7)$$
and consider the subspace ${\cal C}_V\subset {\cal K}_c$ given by
$$
{\cal C}_V := \{ J \,\vert\ J=C_0 + j_V \,,\quad j_V: S^1\to V\,\}.
\eqno(4.8)$$
The assumption is that ${\cal C}_V$ defines a global gauge fixing
in such a way that the components of the gauge fixed current $j_V$,
when considered as functions on ${\cal K}_c$,
provide a free generating set for ${\cal R}$.
All in all, the above assumptions
say that the main features of  the standard
case are valid for the reduction.
Then the following result may be proven.
\medskip
\noindent
{\bf Theorem 2.} {\it
Under the assumptions described above,
there exists an element $M_-\in \Gamma$ which together
with $H$ and $C_0$ generates an $sl_2$ subalgebra of $G$.
More precisely, there exists an element $M_-\in \Gamma$
such that
Eq.~(1.21) holds with $M_0:=H$  and $M_+:=C_0$.}
\smallskip

The proof  of this result,
and further results, implying for example
that most KM reductions through
conformally invariant first class constraints do not result in a
(freely generated) ${\cal W}$-algebra,
can be found in Refs.~[2], [16].
\bigskip\noindent
{\bf Acknowledgments}
\medskip
This report is based on joint work with J. Harnad,
I. Marshall, L. O'Raifeartaigh, P. Ruelle, I. Tsutsui and
A. Wipf, to whom I wish to express my indebtedness.
\bigskip\noindent
{\bf References}
\medskip
\item{[1]}
L. Feh\'er, J. Harnad and I. Marshall,
{\it Generalized Drinfeld-Sokolov Reductions and KdV-Type Hierarchies},
Montreal preprint, UdeM-LPN-TH-92/103, CRM-1832 (1992).
\item{[2]}
L. Feh\'er,  L. O'Raifeartaigh, P. Ruelle, I. Tsutsui and Wipf,
{\it On Hamiltonian reductions of the Wess-Zumino-Novikov-Witten theories},
{\sl Phys. Rep.} (in press).
\item{[3]}
V. G. Drinfeld and V. V. Sokolov,
{\sl Jour. Sov. Math.} {\bf 30} (1985) 1975;
{\sl Soviet. Math. Dokl.} {\bf 23} (1981) 457.
\item{[4]}
M.~A. Semenov-Tian-Shansky,
{\sl Funct. Anal. Appl.} {\bf 17} (1983) 259.
\item{[5]}
A.~G. Reyman and M.~A. Semenov-Tian-Shansky,
{\sl Phys. Lett.} {\bf A130} (1988) 456.
\item{[6]}
M.~F. De Groot, T.~J. Hollowood and J.~L. Miramontes,
{\sl Commun. Math. Phys.} {\bf 145} (1992) 57;
\item{}
N.~J. Burroughs, M.~F. De Groot, T.~J. Hollowood and J.~L. Miramontes,
{\it Generalized Drinfeld-Sokolov Hierachies II: The Hamiltonian Structures},
Princeton preprint PUTP-1263, IASSN-HEP-91/42 (1991);
{\sl Phys. Lett.} {\bf 277B} (1992) 89.
\item{[7]}
J. Balog, L. Feh\'er,  P. Forg\'acs, L. O'Raifeartaigh and Wipf,
{\sl Ann. Phys.} (NY) {\bf 203} (1990) 76.
\item{[8]}
V.~G. Kac and D.~H. Peterson,
in: {\it Proceedings of Symposium on Anomalies, Geometry and Topology},
eds. W.~A. Bardeen and  A.~R. White
(World Scientific, Singapore, 1985).
\item{[9]}
F. ten Kroode and J. van de Leur,
{\sl Commun. Math. Phys.} {\bf 137} (1991) 67.
\item{[10]}
F. ten Kroode and J. van de Leur,
{\it Bosonic and fermionic realizations of
the affine algebra $\widehat{so}_{2n}$},
Utrecht preprint no. 636 (1991);
{\it Level one representations of the affine Lie algebra $B_n^{(1)}$},
Utrecht preprint no. 698 (1991);
{\it Level one representations of the twisted affine algebras
$A_n^{(2)}$ and $B_n^{(2)}$},
Utrect preprint no. 707 (1992).
\item{[11]}
L.~A. Dickey,
{\it Soliton Equations and Hamiltonian Systems}
(World Scientific, Singapore, 1991).
\item{[12]}
I.~M. Gelfand and L.~A. Dickey,
{\sl Funct. Anal. Appl.} {\bf 11} (1977) 93.
\item{[13]}
Yu.~I. Manin,
{\sl Jour. Sov. Math.} {\bf 11} (1979) 1.
\item{[14]}
G. Wilson,
{\sl Math. Proc. Cambr. Philos. Soc.} {\bf 86} (1979) 131;
{\sl Q. J. Math. Oxford} {\bf 32} (1981) 491.
\item{[15]}
F. A. Bais, T. Tjin, and P. van Driel,
{\sl Nucl. Phys.} {\bf B357} (1991) 632.
\item{[16]}
L. Feh\'er,  L. O'Raifeartaigh, P. Ruelle and I. Tsutsui,
{\sl Phys. Lett.} {\bf 283B} (1992) 243;
{\it Limitations on the Classical Reductions of KM-Algebras to
${\cal W}$-Algebras}, preprint to appear.

\bye